\documentclass[aps,pre,reprint,groupedaddress]{revtex4-2}
\usepackage[dvipdfmx]{graphics}
\usepackage{physics}
\usepackage{amsmath,amssymb}
\usepackage{bm}

\begin{document}


\title{
  Two-dimensional hydrodynamic simulation for synchronized oscillatory flows in two collapsible channels connected in parallel
}


\author{Yuki Araya}
\affiliation{Department of Physics, Chiba University, Chiba 263-8522, Japan}
\author{Hiroaki Ito}
\affiliation{Department of Physics, Chiba University, Chiba 263-8522, Japan}
\author{Hiroyuki Kitahata}
\email[]{kitahata@chiba-u.jp}
\affiliation{Department of Physics, Chiba University, Chiba 263-8522, Japan}


\date{\today}

\begin{abstract}
  We investigated self-sustained oscillation in a collapsible channel, in which a part of one rigid wall is replaced by a thin elastic wall, and synchronization phenomena in the two channels connected in parallel. We performed a two-dimensional hydrodynamic simulation in a pair of collapsible channels which merged into a single channel downstream. The stable synchronization modes depended on the distance between the deformable region and the merging point; only an in-phase mode was stable for the large distance, in-phase and antiphase modes were bistable for the middle distance, and again only an in-phase mode was stable for the small distance. An antiphase mode became stable through the subcritical pitchfork bifurcation by decreasing the distance. Further decreasing the distance, the antiphase mode became unstable through the subcritical Neimark-Sacker bifurcation. We also clarified the distance dependences of the amplitude and frequency for each stable synchronization mode.
\end{abstract}


\maketitle


\section{introduction}
Synchronization is observed in various chemical, biological, and hydrodynamic systems such as chemical oscillation in Belousov-Zhabotinsky reactions\cite{Mikhailov2017}, heartbeat as a collective contraction of cardiomyocytes\cite{Kojima2006}, and cooperatively beating flagellae\cite{Taylor1951, Yang2008}. Synchronization with hydrodynamic interactions has been studied to understand oscillation suppression in the vortex streets\cite{Kaneko2008}, efficient swimming and flying in the collective motion of animals\cite{Verma2018, Floryan2018, Li2020}, etc. It was reported that the beating of flagellae synchronizes in an in-phase mode in the collective motion of sperms through the hydrodynamic interaction described by Stokes flow\cite{Taylor1951, Yang2008}. Two K\'arm\'an vortex streets appearing behind two circular cylinders synchronize in an antiphase mode with interaction through the vortex\cite{Sumner2010, Zhou2016}. In flame oscillators, which induce oscillatory flows of gases driven by buoyancy due to the combustion heat and exhibit synchronization through vortex interaction, an in-phase mode is stable for a small distance between the two flame oscillators and an antiphase mode is stable for a large distance\cite{Kitahata2009, Yang2019}. The hydrodynamic interaction usually cannot be described in a simple form due to the nonlinearity of the Navier-Stokes equation, and it leads to complex synchronization phenomena. There have been some reports that explain these synchronization phenomena with low-dimensional models composed of a set of ordinary differential equations, such as phase equations and complex amplitude equations\cite{Pikovsky2002, Kuramoto1984}.

\begin{figure}[tb]
  \includegraphics{"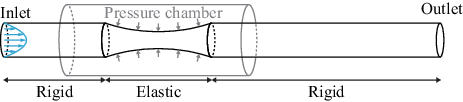"}
  \caption{
    Schematic illustration of a Starling resistor. The elastic tube receives external pressure in a pressure chamber.
  \label{fig:1}} 
\end{figure}

One of the hydrodynamic systems that exhibit nonlinear oscillation is the oscillatory flow in a collapsible tube. It occurs through the interaction between flow and the deformation of an elastic tube that receives external pressure, and thus, is of interest particularly in engineering as a fluid-structure interaction problem\cite{Grotberg2004, Heil2011}. This oscillatory flow in a collapsible tube also attracts interest in physiology since it relates to wheezing during forced expiration\cite{Gavriely1989, Grotberg1989} and Korotkoff sounds during sphygmomanometry\cite{Rodbard1967, Ur1970}. As an ideal model system for the collapsible tube, a Starling resistor has been investigated in the context of fundamental studies for these phenomena\cite{Paidoussis2014}. The Starling resistor is composed of an elastic tube sandwiched by two rigid tubes, as shown in Fig.~\ref{fig:1}. By introducing a steady flow at the inlet and an external pressure around the elastic tube, the oscillatory flow with the periodic deformation of the elastic tube and the periodic vortex shedding downstream of the deformable region is observed. An axisymmetric steady flow in an elastic tube bifurcates into a non-axisymmetric steady flow in a buckled tube with an increase in the external pressure, and with further increasing the external pressure, the steady flow in the buckled tube bifurcates into an oscillatory flow. Previous studies from the viewpoint of the dynamical systems investigated whether the system has hysteresis near the bifurcation point from an axisymmetric state to a buckled state when the system parameters are changed\cite{Flaherty1972, Bertram1987, Bertram1989, Heil1996, Heil2010, Kozlovsky2014, Gregory2021, Laudato2023, Zhang2023}. The bifurcation phenomena from the stationary state to the oscillatory state were theoretically studied using the collapsible channel\cite{Pedley1992}, which is the two-dimensional hydrodynamic model shown in Fig.~\ref{fig:2}(a). The bifurcation points where supercritical Hopf bifurcation occurs were identified for various parameters such as a Reynolds number, stretching stiffness, and external pressure\cite{Luo2008, Liu2012, Hao2016, Wang2021, Wang2021_2}. The entrainment phenomena, i.e. the frequency locking at an integer ratio between external forcing and oscillation frequencies, were observed in the self-sustained oscillatory flow with periodic pressure modulation in the upstream\cite{Bertram1992, She1996, Bertram2000, Bertram2000a} or pressure chamber\cite{Kumar2022}.

Since the entrainment phenomena were observed in the single collapsible tube with external forcing, the connected collapsible tubes are expected to exhibit synchronization phenomena. Oscillatory flow in a collapsible tube or channel shows large pressure changes due to not only vortex shedding but also the moving boundary. Thus, the coupled system is expected to exhibit novel synchronization phenomena, which may not be described by the phase reduction method. To investigate the synchronization phenomena of the oscillatory flows in two coupled systems, we consider two collapsible channels connected in parallel, in which two channels merged into a single channel, as shown in Fig.~\ref{fig:2}(b).

In this paper, we investigate the oscillatory flow in two collapsible channels connected in parallel from the viewpoint of synchronization phenomena and clarify the nonlinear dynamics of this system. We performed a two-dimensional numerical simulation varying the distance between the deformable region and the merging point, and identify the distance dependences of the stable synchronization mode, amplitude, and frequency. We solve the governing equation of incompressible viscous fluid with the lattice Boltzmann method\cite{Chen1998, Kruger2017, Inamuro2021}. We employ a modified model from the one proposed by Wang {\it et al.}\cite{Wang2021} as the governing equation of the elastic wall. For fluid-structure interaction, we adopt the immersed boundary method based on the one used by Huang {\it et al.}\cite{Huang2021}. We describe the hydrodynamic model for the collapsible channel in Sec.~II and the numerical method to solve the governing equations in Sec.~III. Section~IV shows the results of a single collapsible channel to validate our simulation model and calculation method. Then, Sec.~V shows the results of the synchronization in the two collapsible channels connected in parallel. Section~VI mainly discusses the stability of the synchronization mode from the viewpoint of phase dynamics. Finally, Sec.~VII concludes the obtained results on the synchronization of the two collapsible channels connected in parallel.

\section{model}
\begin{figure}[tb]
   \includegraphics{"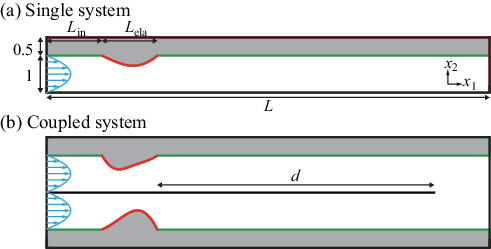"}
  \caption{
    Schematic illustrations of the two-dimensional model of (a)~a single collapsible channel and (b)~two coupled collapsible channels. The red and green lines indicate the immersed boundary of the elastic and rigid walls, respectively. The gray regions indicate the virtual fluid. The Poiseuille flow at the inlet (left side) is indicated by the blue lines and arrows.
  \label{fig:2}}
\end{figure}

We consider a two-dimensional incompressible viscous flow in a channel, where a part of one rigid wall is replaced by an elastic wall, as shown in Fig.~\ref{fig:2}. The non-dimensionalized governing equations for the velocity field are the equation of continuity and the Navier-Stokes equation, given as
\begin{align}
  &\pdv{u_\alpha}{x_\alpha}=0,
\end{align}
\begin{align}
  &\pdv{u_\alpha}{t}+u_\beta\pdv{u_\alpha}{x_\beta}=\pdv{\sigma_{\alpha\beta}}{x_\beta},
\end{align}
\begin{align}
  &\sigma_{\alpha\beta}=-P\delta_{\alpha\beta}+\frac{1}{\rm Re}\qty(\pdv{u_\alpha}{x_\beta}+\pdv{u_\beta}{x_\alpha}),\label{eq:7}
\end{align}
where \(t\) is the time, \(x_\alpha(\alpha=1,2)\) is the spatial coordinate, \(u_\alpha\) is the velocity field, \(P\) is the pressure field, and \(\sigma_{\alpha\beta}\) is the hydrodynamic stress. We use the subscripts \(\alpha\) and \(\beta\) to denote the spatial coordinates and adopt the summation convention. \(\delta_{\alpha\beta}\) is the Kronecker delta. \(\rm Re\) is the Reynolds number, which is the ratio of the inertial force to the viscous one. We employ the average velocity at the inlet, the channel width, and fluid density to the characteristic velocity, length, and density scales, respectively. All the variables and parameters are non-dimensionalized by them.

We consider the deformable wall, whose thickness is infinitesimally small. We introduce not only the elasticity but also the small viscosity into the deformable wall to stabilize the numerical simulation. We set the viscosity small enough to approximate the wall as an elastic wall and hereafter call it an elastic wall. We set the initial position of the elastic wall such that the wall is parallel to the \(x_1\)-axis and exerts no elastic force. We denote the position vector of the deformed elastic wall as \(X_\alpha(l)\) and derivative with respect to \(l\) as \({}'\), where \(l\) is the initial \(x_1\)-coordinate of the elastic wall. The elastic energy per unit length is composed of the stretching energy \(K_{\rm s}(s-1)^2/2\) and bending energy \(K_{\rm b}\kappa^2/2\), where \(s(l)=\sqrt{X_1'^2+X_2'^2}\) is the stretching deformation, \(\kappa(l)=(X_2''X_1'-X_2'X_1'')/s^2\) is the curvature of the wall, \(K_{\rm s}\) is the stretching stiffness, and \(K_{\rm b}\) is the bending stiffness\cite{Wang2021}. Note that the elastic energy is zero in the initial position. The viscous energy dissipation per unit length is induced by the stretching deformation velocity as \(\nu(v_\alpha'\tau_\alpha)^2\), where \(v_\alpha(l)=\partial X_\alpha/\partial t\) is the velocity, \(\tau_\alpha(l)\) is the unit tangential vector of the elastic wall, and \(\nu\) is the viscosity of the elastic wall. The conservation law of energy per unit length is expressed as
\begin{align}
  \pdv{t}\qty(\frac{M}{2}v_\alpha^2+\frac{K_{\rm s}}{2}(s-1)^2+\frac{K_{\rm b}}{2}\kappa^2)=-\nu(v_\alpha'\tau_\alpha)^2,
\end{align}
where \(M\) is the mass density ratio of the elastic wall to the fluid. We add external force terms considering the existence of fluids on both sides of the elastic wall, and we obtain the equation of motion for the elastic wall by performing the time derivative and integration by parts with respect to \(l\), as
\begin{align}
  M\pdv{v_\alpha}{t}=&\qty(K_{\rm s}(s-1)\tau_\alpha-K_{\rm b}\frac{\kappa'}{s}n_\alpha+\nu v_\beta'\tau_\beta\tau_\alpha)'\nonumber\\
  &+s(\sigma_{\alpha\beta}^+-\sigma_{\alpha\beta}^-)n_\beta,\label{eq:8}
\end{align}
where \(n_\alpha\) is the unit normal vector to the elastic wall. \(\sigma_{\alpha\beta}^+,\) and \(\sigma_{\alpha\beta}^-\) are the hydrodynamic stresses exerted on the wall from the up-side and down-side fluids, respectively.

For a single collapsible channel, the channel width is 1 and the length is \(L\). A part of the upper wall is replaced by an elastic wall, whose length is \(L_{\rm ela}\), as shown in Fig.~\ref{fig:2}(a). In this situation, the elastic wall receives the hydrodynamic stress \(\sigma_{\alpha\beta}^-\) obtained by eq.~\eqref{eq:7} from the lower surface, and the external stress \(\sigma_{\alpha\beta}^+=-P_{\rm ext}\delta_{\alpha\beta}\) from the upper surface, where \(P_{\rm ext}\) is the constant pressure. We set the inlet at \(x_1=0\) and the outlet at \(x_1=L\). Three rigid walls are set at \(\{(x_1,x_2)|\, 0\leq x_1\leq L, x_2=0\}\), \(\{(x_1,x_2)|\, 0\leq x_1\leq L_{\rm in}, x_2=1\}\), and \(\{(x_1,x_2)|\, L_{\rm in}+L_{\rm ela}\leq x_1\leq L, x_2=1\}\). We insert an elastic wall at \(\{(x_1,x_2)|\, L_{\rm in}<x_1<L_{\rm in}+L_{\rm ela}, x_2=1\}\) initially. We set \(L=40\) and \(L_{\rm in}=L_{\rm ela}=5\), following the previous study\cite{Wang2021}. At the inlet (\(x_1=0\)), we apply the Poiseuille flow, where the direction is along the \(x_1\)-axis and the mean speed is equal to 1, expressed as \(u_1=6x_2\qty(1-x_2),u_2=0\). At the outlet (\(x_1=L\)), fluid can flow out under a constant pressure, \(P=0\). We set the boundary condition for the elastic wall \(X_\alpha={\rm const.}\) and \(X_\alpha''=0\) at \(l=L_{\rm in}\) and \(l=L_{\rm in}+L_{\rm ela}\). We vary \(K_{\rm s}\) and \(P_{\rm ext}\) in the range of \(500\leq K_{\rm s}\leq3000\) and \(1.6\leq P_{\rm ext}\leq3.0\). We set \({\rm Re}=300,K_{\rm b}=10^{-5}K_{\rm s}\), and \(\nu=0.25\). In the initial condition at \(t=0\), the elastic wall is flat, and the flow profile at every \(x_1\) is the same as that at the inlet \(x_1=0\). The pressure distribution has a linear gradient along the \(x_1\)-axis and is constant along the \(x_2\)-axis to be consistent with the profile of the Poiseuille flow. We perform numerical simulation within the time range of \(0\leq t\leq250\).

For two coupled collapsible channels, we employ the same geometry as the single collapsible channel for \(x_2\geq 0\) and the mirror inversion of it with respect to \(x_2=0\) for \(x_2\leq0\), and remove the wall at \(\{(x_1,x_2)|\, L_{\rm in}+L_{\rm ela}+d\leq x_1\leq L, x_2=0\}\) to couple them, as shown in Fig.~\ref{fig:2}(b). \(d\) is the distance between the right end of the deformable region and the merging point and is varied in the range of \(20\leq d\leq 30\). Note that there is no interaction at \(d=30\). We fix \(K_{\rm s}=1000\) and \(P_{\rm ext}=3.0\), and set the other parameters to be the same as those in the case with the single collapsible channel. The initial conditions at \(t=-250\) for the coupled system are also the same as those for the single collapsible channel. First, we calculate without the interaction, {\it i.e.}, without removing the wall at \(x_2=0\), and without the deformation of both elastic walls (\(v_\alpha=0\)) in the time range of \(-250\leq t<-200\). Then, we allow the upper elastic wall (\(x_2>0\)) to move after \(t=-200\) and proceed with numerical simulation without the interaction till \(t=0\). To introduce the delay in phase, we make the lower elastic wall (\(x_2<0\)) start to move after \(t=-200+T\), where \(T\) is the time delay, without the interaction till \(t=0\). We remove the wall at \(x_2=0\) for \(L_{\rm in}+L_{\rm ela}+d\leq x_1\leq L\) at \(t=0\) and calculate with the interaction till \(t=250\). \(T\) is varied from 0 to 2.975 (period of a single oscillator) in increments of 0.0875, which corresponds to varying the initial phase difference between the two oscillators.

\section{numerical method}
We adopt the D2Q9 lattice Boltzmann method with a multi-relaxation-time model for the two-dimensional hydrodynamics\cite{DeRosis2017, Fei2017}. In this method, we consider the nine discrete velocities \(c_{k\alpha}\,(k=0,1,\dots,8)\) and the nine corresponding particle distribution functions \(f_k(x_1,x_2,t)\). We denote the set of nine distribution functions as a vector \(\bm f\). We can obtain the hydrodynamic velocity as \(u_\alpha=\qty(\sum_kf_kc_{k\alpha})/\qty(\sum_kf_k)\) when \(\bm f\) follows the discrete Boltzmann equation,
\begin{align}
  f_k(x_1+c_{k1}\Delta x,x_2+c_{k2}\Delta x,t+\Delta t)-f_k(x_1,x_2,t)\nonumber\\
  =\Gamma_k(\bm f(x_1,x_2,t)),
\end{align}
where \(\Delta t\) is the time step, \(\Delta x\) is the spatial mesh, and \(\Gamma_k\) is the collision term for each \(f_k\). In the multi-relaxation-time model, the collision term is expressed as
\begin{align}
  \bm \Gamma=&{\sf M}^{-1}{\sf N}^{-1}{\sf S}{\sf N}{\sf M}(\bm f(x_1,x_2,t)-\bm f^{\rm eq}(x_1,x_2,t)),
\end{align}
where \(\bm \Gamma\) is the vector composed of the nine collision terms, \(\sf M\) is ``transformation matrix'', \(\sf N\) is ``shift matrix'', and \(\sf S\) is ``diagonal relaxation matrix''. \({\sf M}^{-1}\) and \({\sf N}^{-1}\) are the inverse matrices of \(\sf M\) and \(\sf N\), respectively, and \(\bm f^{\rm eq}\) is the equilibrium distribution function\cite{DeRosis2017,Fei2017}. The equilibrium distribution function is given as,
\begin{align}
  f^{\rm eq}_k=\rho\omega_k\qty(1+3c_{k\alpha}u_\alpha-\frac{3}{2}u_\alpha u_\alpha+\frac{9}{2}(c_{k\alpha}u_\alpha)^2),
\end{align}
where the weight \(\omega_k\) is set to be
\begin{align}
  \omega_k=\left\{\begin{array}{ll}
    4/9&(k=0)\\
    1/9&(k=1,2,3,4)\\
    1/36&(k=5,6,7,8)
  \end{array}\right.,
\end{align}
the nine discrete velocities are set to
\begin{align}
    &c_{0\alpha}=(0,0),\\
    &c_{1\alpha}=(1,0),\\
    &c_{2\alpha}=(0,1),\\
    &c_{3\alpha}=(-1,0),\\
    &c_{4\alpha}=(0,-1),\\
    &c_{5\alpha}=(1,1),\\
    &c_{6\alpha}=(-1,1),\\
    &c_{7\alpha}=(-1,-1),\\
    &c_{8\alpha}=(1,-1),
\end{align}
and \(\rho=\sum_kf_k\) is the pseudo-density. \(\sf M\) transforms \(\bm f\) to the ``raw moment'', as
\begin{align}
  {\sf M}\bm f=\sum_k\mqty(
    f_k\\
    f_kc_{k1}\\
    f_kc_{k2}\\
    f_k({c_{k1}}^2+{c_{k2}}^2)\\
    f_k({c_{k1}}^2-{c_{k2}}^2)\\
    f_kc_{k1}c_{k2}\\
    f_k{c_{k1}}^2c_{k2}\\
    f_kc_{k1}{c_{k2}}^2\\
    f_k{c_{k1}}^2{c_{k2}}^2
  ),
\end{align}
and \(\sf N\) transforms \({\sf M}\bm f\) to ``central moment'', as
\begin{align}
  {\sf NM}\bm f=\sum_k\mqty(
    f_k\\
    f_k(c_{k1}-u_1)\\
    f_k(c_{k2}-u_2)\\
    f_k((c_{k1}-u_1)^2+(c_{k2}-u_2)^2)\\
    f_k((c_{k1}-u_1)^2-(c_{k2}-u_2)^2)\\
    f_k(c_{k1}-u_1)(c_{k2}-u_2)\\
    f_k(c_{k1}-u_1)^2(c_{k2}-u_2)\\
    f_k(c_{k1}-u_1)(c_{k2}-u_2)^2\\
    f_k(c_{k1}-u_1)^2(c_{k2}-u_2)^2\\
  ).
\end{align}
We use \({\sf S}={\rm diag}(0,0,0,0.5,1/(0.5+3/{\rm Re}),1/(0.5+3/{\rm Re}),0.8,0.8,0.5)\) for the numerical stability.

We adopt the finite difference scheme to solve eq.~\eqref{eq:8} for the elastic wall. We handle the spatial and time derivative terms with the central difference scheme. We employ the same spatial mesh, \(\Delta x\), as used in the lattice Boltzmann method.

For the coupling between fluid and structure dynamics, we adopt the immersed boundary method\cite{Peskin2002} for thin elastic and rigid walls. Hereafter, we denote the initial position, deformed position, and the velocity not only for the elastic wall but also for the rigid wall as \(l\), \(X_\alpha\), and \(v_\alpha\), respectively. Note that they follows \(X_1=l\), \(X_2={\rm const.}\) and \(v_\alpha=0\) for the rigid wall. According to the law of action and reaction, the fluid on both sides of a thin wall follows
\begin{align}
  \pdv{u_\alpha}{t}+u_\beta\pdv{u_\alpha}{x_\beta}=&\pdv{\sigma_{\alpha\beta}}{x_\beta}+(\sigma_{\alpha\beta}^--\sigma_{\alpha\beta}^+)n_\beta\Phi,
\end{align}
where \(\Phi(\cdot)\) represents the Dirac delta function. Hereafter, the argument of \(\Phi\) is \(\sqrt{(x_\gamma-X_\gamma)(x_\gamma-X_\gamma)}\) in the case it is omitted. Note that the non-slip boundary condition at the thin wall is expressed by the stress from the thin wall in this equation. Using the fractional step method, the fluid follows
\begin{align}
  u_\alpha^{*(0)}=u_\alpha(t)+\qty(-u_\beta\pdv{u_\alpha}{x_\beta}+\pdv{\sigma_{\alpha\beta}}{x_\beta})\Delta t,
\end{align}
\begin{align}
  u_\alpha(t+\Delta t)=u_\alpha^{*(0)}+(\sigma_{\alpha\beta}^--\sigma_{\alpha\beta}^+)n_\beta \Phi\Delta t.\label{eq:2}
\end{align}
Since the velocity of the fluid at the thin wall should correspond to that of the thin wall in the discrete space, the stress is expressed as
\begin{align}
  (\sigma_{\alpha\beta}^--\sigma_{\alpha\beta}^+)n_\beta=\frac{v_\alpha(t+\Delta t)-u_\alpha^{*(0)}}{\Delta t}\Delta x.\label{eq:4}
\end{align}
With the spatial discretization, we approximate the Dirac delta function with
\begin{align}
  &\Phi(\sqrt{x_\alpha x_\alpha})=\frac{1}{\Delta x}\phi(x_1)\phi(x_2),
\end{align}
\begin{align}
  &\phi(x)=\left\{\begin{array}{ll}
    \qty(\cos\qty(\pi x/(2\Delta x))+1)/4&(\qty|x|<2\Delta x)\\
    0&(\qty|x|\geq2\Delta x)\\
  \end{array}\right.,
\end{align}
and the complement of the velocity on the thin wall is expressed as
\begin{align}
  \tilde u_\alpha^{*(0)}(l)=&\frac{1}{\Delta x}\int u_\alpha^{*(0)}(x_1,x_2)\Phi\,\dd x_1\dd x_2.\label{eq:3}
\end{align}
We adopt the simple forcing scheme\cite{He1997} to calculate eq.~\eqref{eq:2} in the lattice Boltzmann method.

Instead of eqs.~\eqref{eq:2} and \eqref{eq:3}, the iteration in a single time step
\begin{align}
  u_\alpha^{*(m+1)}(x_1,x_2)=&u_\alpha^{*(m)}(x_1,x_2)\nonumber\\
  &+\qty(v_\alpha(l)-\tilde u_\alpha^{*(m)}(l))\Phi\Delta x,
\end{align}
\begin{align}
  \tilde u_\alpha^{*(m)}(l)=&\frac{1}{\Delta x}\int u_\alpha^{*(m)}(x_1,x_2)\Phi\,\dd x_1\dd x_2,
\end{align}
where \(m\) is a non-negative integer and \(u_\alpha^{*(m)}\) is the fluid velocity obtained after \(m\) times iteration, can improve the numerical accuracy\cite{Wang2008, Suzuki2011}. We perform the iteration five times in each time step and employ \(u_\alpha^{*(5)}\) as \(u_\alpha(t+\Delta t)\) instead of the right-hand side of eq.~\eqref{eq:2}.

In a collapsible channel, the elastic wall receives the hydrodynamic stress written in eq.~\eqref{eq:7} only on one side and receives the constant pressure \(P_{\rm ext}\) on the other side. To adopt the immersed boundary method for the collapsible channel, we assume a virtual fluid, which receives the stress from the elastic wall and does not apply any stress to the elastic wall, on the other side of the actual fluid. We need to obtain \(\sigma_{\alpha\beta}^-n_\beta\) in the case the virtual fluid is on the upside of the elastic wall. If we know the mean of \(\sigma_{\alpha\beta}^+\) and \(\sigma_{\alpha\beta}^-\) as \(\sigma_{\alpha\beta}^{\rm avg}\), \(\sigma_{\alpha\beta}^-n_\beta\) is obtained as
\begin{align}
  \sigma_{\alpha\beta}^-n_\beta=\sigma_{\alpha\beta}^{\rm avg} n_\beta+\frac{\sigma_{\alpha\beta}^--\sigma_{\alpha\beta}^+}{2}n_\beta,
\end{align}
where the second term on the right-hand side is derived from eq.~\eqref{eq:4}. We assume that \(\sigma_{\alpha\beta}^{\rm avg}\) is given as
\begin{align}
  \sigma_{\alpha\beta}^{\rm avg}(l)=\frac{1}{\Delta x}\int\sigma_{\alpha\beta}(x_1,x_2)\Phi\,\dd x_1\dd x_2.\label{eq:9}
\end{align}
To calculate the integral in eq.~\eqref{eq:9}, we obtain the hydrodynamic stress \(\sigma_{\alpha\beta}\) from the lattice Boltzmann method as
\begin{align}
  \sigma_{\alpha\beta}=&-\frac{1}{2\tau}\sum_k\frac{1}{3}f_k\delta_{\alpha\beta}+\frac{1}{3}\delta_{\alpha\beta}\nonumber\\
  &-\qty(1-\frac{1}{2\tau})\sum_kf_k(c_{k\alpha}-u_\alpha)(c_{k\beta}-u_\beta),
\end{align}
where \(\tau=0.5+3/{\rm Re}\).

For a single collapsible channel, we consider a rectangular region, \(0\leq x_1\leq L,0\leq x_2\leq 1.5\). We adopt the Dirichlet boundary condition, \(u_1=6x_2\qty(1-x_2),u_2=0\), for \(x_1=0,0\leq x_2\leq1\), non-slip boundary condition, \(u_1=u_2=0\), for \(x_1=0,1<x_2\leq1.5\), and the Neumann boundary condition, \(P=0,u_2=0\), for \(x_1=L,1<x_2\leq1.5\), using the on-grid bounce-back scheme\cite{Zou1997}. The rigid walls at \(x_2=0\) and \(1.5\) are handled with the standard half-way bounce-back scheme. We impose the immersed boundary at \(x_2=1\) for the initial condition. We set \(\Delta t=2.5\times10^{-4}\) and \(\Delta x=2.5\times10^{-2}\). For the two coupled collapsible channels, we set the same conditions as those for the single collapsible channel for \(x_2\geq 0\) and the mirror inversion of it with respect to \(x_2=0\) for \(x_2\leq0\).

\section{single collapsible channel}
\begin{figure*}[tb]
  \includegraphics{"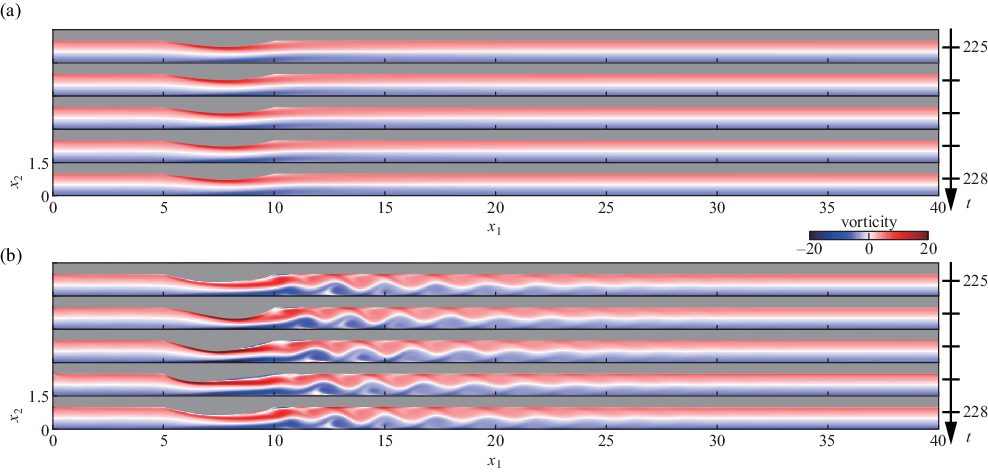"}
  \caption{
    Sequential snapshots of the vorticity field for (a)~\(K_{\rm s}=1000,P_{\rm ext}=1.6\) and (b)~\(K_{\rm s}=1000,P_{\rm ext}=3.0\) after a sufficiently long time \(225\leq t\leq228\) from the initial state.
  \label{fig:3}}
\end{figure*}

The sequential snapshots of the vorticity field \(\partial u_2/\partial x_1-\partial u_1/\partial x_2\) in a single channel after a sufficiently long time from the initial state are shown in Fig.~\ref{fig:3}. Figure~\ref{fig:3}(a) shows the stationary state, in which the elastic wall does not move, obtained for \(K_{\rm s}=1000\) and \(P_{\rm ext}=1.6\). Figure~\ref{fig:3}(b) shows the oscillatory state, in which the deformation of the elastic wall and the vortex shedding downstream of the deformable region are periodic in time, obtained for \(K_{\rm s}=1000\) and \(P_{\rm ext}=3.0\).

\begin{figure}[tb]
  \includegraphics{"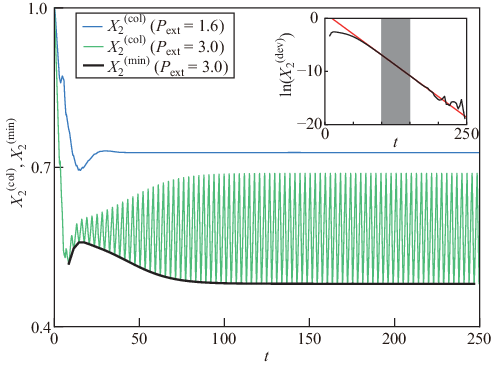"}
  \caption{
    Time series of \(X_2^\qty(\rm col)\), the spatial minimum value of \(X_2\). The blue and green lines correspond to Figs.~\ref{fig:3}(a) and (b), respectively. The black line indicates the time series of \(X_2^\qty(\rm min)\), which is the local minimum of \(X_2^\qty(\rm col)\) with respect to the time. The natural logarithmic plot of \(X_2^\qty(\rm dev)\), which is the deviation of \(X_2^\qty(\rm min)(t)\) from the converged value of \(X_2^\qty(\rm min)\), is shown in the inset. The red line shows the result of linear fitting using the time range of \(100\leq t\leq150\), indicated by the shaded region.
  \label{fig:4}}
\end{figure}

To investigate the time series of the deformation, we define the spatial minimum value of \(x_2\) coordinates of the elastic wall as \(X_2^\qty(\rm col)(t)=\min_{5\leq l\leq10}|X_2(t,l)|\). The time series of \(X_2^\qty(\rm col)\) corresponding to Figs.~\ref{fig:3}(a) and \ref{fig:3}(b) are shown in Fig.~\ref{fig:4}. \(X_2^\qty(\rm col)\) converged to a constant value in the stationary state. The oscillation amplitude, which is defined by the difference between the local maximum and minimum values of \(X_2^\qty(\rm col)\) with respect to time, converged to a finite constant value in the oscillatory state. Note that the maximum discrepancies between the result at \(\nu=0.25\) and that at \(\nu=0.5\) or 0.125 were less than 0.2\%, and thus the viscosity of the wall is small enough to approximate that the wall is elastic. From the viewpoint of continuous dynamical systems, the stationary and oscillatory states correspond to the stable fixed point and the stable limit-cycle oscillation, respectively. To investigate the dynamics of the convergence to the limit cycle, we measured \(X_2^\qty(\rm min)\), the local minimum values of \(X_2^\qty(\rm col)\) with respect to time, and \(X_2^\qty(\rm dev)\), the deviation of \(X_2^\qty(\rm min)\) from the converged value of \(X_2^\qty(\rm min)\), in the oscillatory state. We employed \(X_2^\qty(\rm min)\) around \(t=250\) as the converged value of \(X_2^\qty(\rm min)\). \(X_2^\qty(\rm dev)\) was exponentially converged after \(t=50\) as shown in the inset of Fig.~\ref{fig:4}. We obtained the linear fitting line, where the slope was \(-0.078\), in the logarithmic plot using the data within the time range of \(100\leq t\leq150\). Note that this slope corresponds to the second maximum Lyapunov exponent and will be used for the discussion of the interaction strength in Sec.~VI.

\begin{figure}[tb]
   \includegraphics{"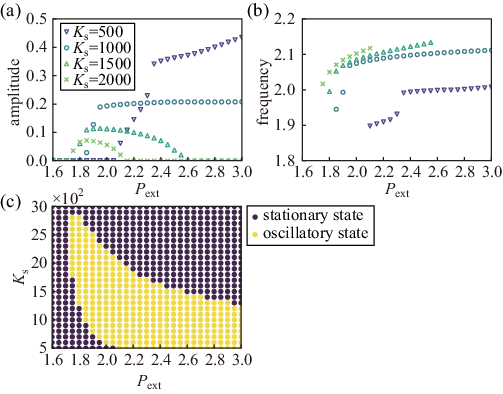"}
  \caption{
    (a)~Amplitude and (b)~frequency of \(X_2^\qty(\rm col)\) depending on \(P_{\rm ext}\) for various \(K_{\rm s}\). The same symbols are used for the legends in panels~(a) and (b). (c)~Phase diagram to distinguish the stationary and oscillatory states in the \(P_{\rm ext}\)-\(K_{\rm s}\) plane.
  \label{fig:5}}
\end{figure}

In the oscillatory state, we employ the time duration between the local minima of \(X_2^\qty(\rm col)\) with respect to time as the period. Then, the (angular) frequency is calculated as \(2\pi\) divided by the period. Note that the amplitude is zero in the stationary state. Figures~\ref{fig:5}(a) and \ref{fig:5}(b) show the \(P_{\rm ext}\)-dependences of the amplitude and frequency. The phase diagram on the \(P_{\rm ext}\)-\(K_{\rm s}\) plane is shown in Fig.~\ref{fig:5}(c). The stationary state bifurcated into the oscillatory state with increasing \(P_{\rm ext}\) for \(K_{\rm s}\leq2800\). The system exhibits reentrance to the stationary state with increasing \(P_{\rm ext}\) for \(1300\leq K_{\rm s}\leq2800\). The stationary state was stable for \(K_{\rm s}\geq2900\). The frequency was increased as \(P_{\rm ext}\) or \(K_{\rm s}\) increased.

\section{two collapsible channels connected in parallel}
\begin{figure*}[tb]
  \includegraphics{"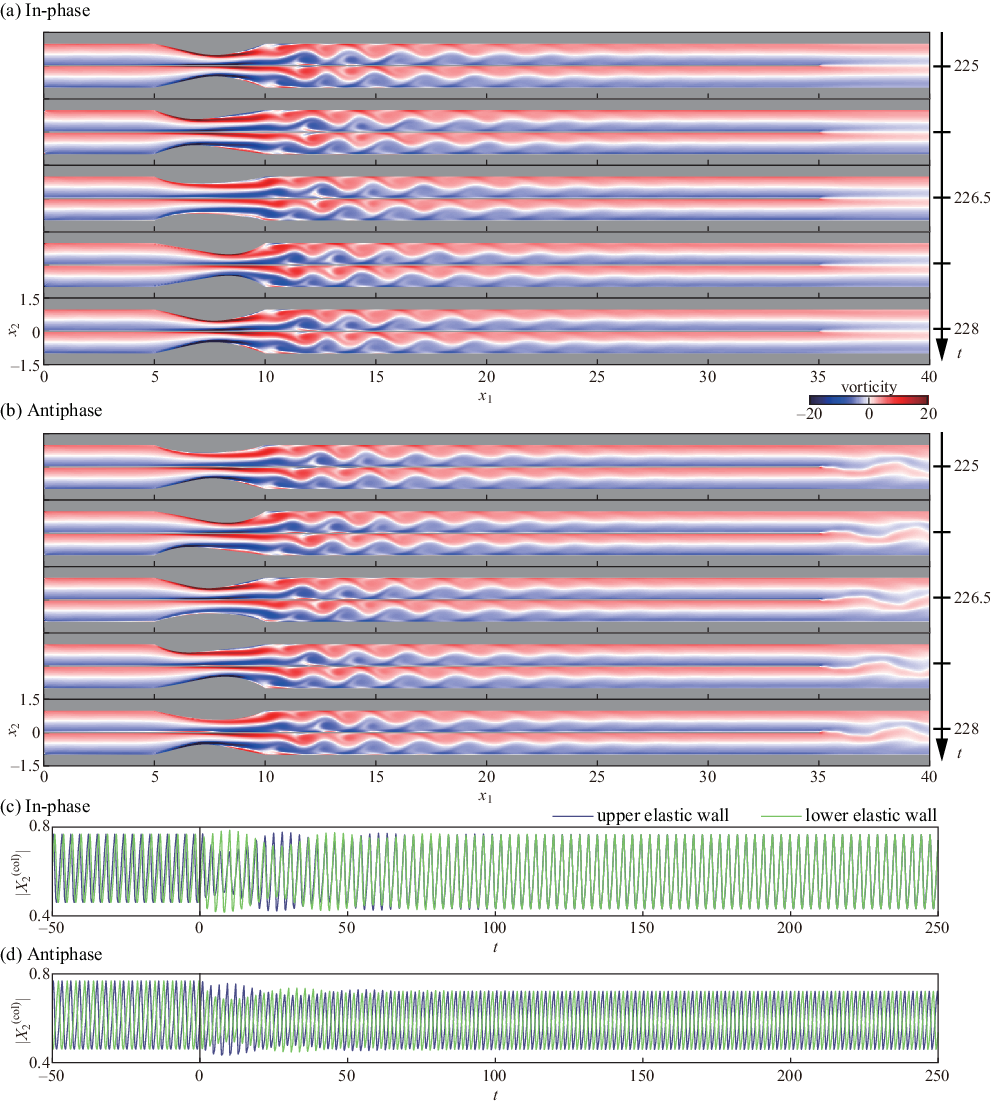"}
  \caption{
    Sequential snapshots of the vorticity field for \(d=25\) in the (a)~in-phase and (b)~antiphase modes after a sufficiently long time \(225\leq t\leq228\) from the state at \(t=0\). (c, d)~Time series of \(|X_2^{(\rm col)}|\) in (c)~in-phase and (d)~antiphase modes, which correspond to panels (a) and (b), respectively.
  \label{fig:6}}
\end{figure*}

The sequential snapshots of the vorticity field in two collapsible channels connected in parallel after a sufficiently long time and the time series of \(\qty|X_2^\qty(\rm col)|\) are shown in Fig.~\ref{fig:6}. The in-phase synchronization mode, in which the elastic walls deform simultaneously, is shown in Figs.~\ref{fig:6}(a) and \ref{fig:6}(c) for \(d=25, T=0.525\). The antiphase synchronization mode, in which the elastic walls deform alternately, is shown in Figs.~\ref{fig:6}(b) and \ref{fig:6}(d) for \(d=25, T=1.75\).

\begin{figure}[tb]
  \includegraphics{"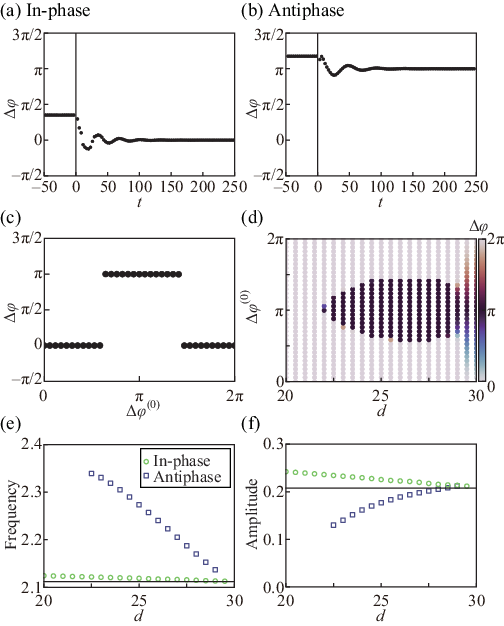"}
  \caption{
    Time series of the phase difference in the (a)~in-phase and (b)~antiphase synchronization modes, which correspond to Fig. \ref{fig:6}(a) for \(\Delta\varphi^\qty(0)=6\pi/17\) and Fig. \ref{fig:6}(b) for \(\Delta\varphi^\qty(0)=20\pi/17\), respectively. (c)~Phase difference at \(t=250\) depending on the initial phase difference \(\Delta\varphi^\qty(0)\) for \(d=25\). (d)~Phase difference at \(t=250\) depending on \(\Delta\varphi^\qty(0)\) and \(d\). (e)~Frequency and (f)~amplitude depending on \(d\) for the in-phase and antiphase synchronization modes. The same symbols are used for the legends in panels~(e) and (f). The horizontal lines show the frequency and amplitude with \(d=30\), which corresponds to the case without interaction.
  \label{fig:7}}
\end{figure}

We regard each of the oscillatory flows in the upper and lower collapsible channels as a limit-cycle oscillator and also regard these oscillatory flows, which interact through the connection, as coupled oscillators. We define a phase \(\varphi\) in each oscillator. We set \(\varphi=0\) when \(X_2^\qty(\rm col)\) has the local minimum value with respect to time. \(\varphi\) increases proportionally to time in an isolated oscillator and is affected by the interaction in the two coupled oscillators. The phase difference between the two oscillators is represented as \(\Delta\varphi=\varphi_{\rm U}-\varphi_{\rm L}\), where \(\varphi_{\rm U}\) and \(\varphi_{\rm L}\) are the phases of the oscillatory flow in the upper and lower collapsible channels, respectively. By varying \(T\) from 0 to 2.975 in increments of 0.0875, the initial phase difference of the oscillators, \(\Delta\varphi^\qty(0)\), was controlled from 0 to \(2\pi\) in increments of \(\pi/17\). The time series of \(\Delta\varphi\) for the parameters in Figs.~\ref{fig:6}(a) and \ref{fig:6}(b) are shown in Figs.~\ref{fig:7}(a) and \ref{fig:7}(b), respectively. The phase difference converged to zero in the in-phase synchronization mode as shown in Fig.~\ref{fig:7}(a) and converged to \(\pi\) in the antiphase synchronization mode as shown in Fig.~\ref{fig:7}(b). The \(\Delta\varphi^\qty(0)\)-dependence of \(\Delta\varphi\) after a sufficiently long time is shown in Fig.~\ref{fig:7}(c) for \(d=25\). Since the final states depended on the initial phase difference, the in-phase and antiphase modes were bistable. The phase difference \(\Delta\varphi\) at \(t=250\) for various \(\Delta\varphi^\qty(0)\) and \(d\) is summarized in Fig.~\ref{fig:7}(d). The in-phase and antiphase modes were clearly bistable for \(22.5\leq d\leq28.5\). The in-phase mode was stable and the antiphase mode was unstable for \(d\leq21.5\). The frequency in each mode is shown in Fig.~\ref{fig:7}(e). The frequency in the antiphase mode was greater than the intrinsic frequency, and the frequency in the in-phase mode was slightly greater than the intrinsic frequency, which is the one for a single collapsible channel with the same parameters. The frequency was decreased with an increase in \(d\) in both the in-phase and antiphase modes. Next, the amplitude in each mode is shown in Fig.~\ref{fig:7}(f). The amplitude in the antiphase mode was increased with an increase in \(d\). The amplitude in the in-phase mode was greater than the intrinsic amplitude and was decreased with an increase in \(d\) in both the in-phase and antiphase modes.

\section{discussion}
We investigated \(P_{\rm ext}\)- and \(K_{\rm s}\)-dependences of the stable state in a single collapsible channel as shown in Fig.~\ref{fig:5}(c). The previous study by Hao {\it et al.}\cite{Hao2016} reported the phase diagram of the stable state on the \(\rm Re\)-\(K_{\rm s}\) plane by linear stability analysis assisted by numerical calculation. In their results for \({\rm Re}=300\) and \(P_{\rm ext}=1.95\), the stationary state bifurcates into the oscillatory state with increasing \(K_s\) and the system exhibits reentrance to the stationary state with further increasing \(K_s\), where the two bifurcation points are \(K_s=447\) and \(K_s=1937\). In our results, there were two bifurcation points but the values were slightly different from those in the previous study: one is in the range between 600 and 700 and the other is in the range between 2300 and 2400, for \({\rm Re}=300\) and \(P_{\rm ext}=1.95\). These discrepancies may be due to the differences in model settings and/or numerical methods. The magnitude of these discrepancies is small, and thus, our model, where we applied the viscosity and mass in the elastic wall for numerical stability, provides reliable results, such as \(P_{\rm ext}\)-dependence and the synchronization modes in the coupled system. The phase diagram on the \(P_{\rm ext}\)-\(K_{\rm s}\) plane was qualitatively consistent with that on the \(\rm Re\)-\(K_{\rm s}\) plane reported by Hao et al.\cite{Hao2016}. In our calculation, the pressure field has an almost linear gradient along the \(x_1\)-axis in the stationary state as shown in Appendix~\ref{app:1}. In the Poiseuille flow with a fixed inflow and outlet pressure, the slope of the linear gradient in the pressure field should approach zero with an increase in \(\rm Re\) by reducing viscosity. Thus, the effect on the elastic wall by increasing \(\rm Re\) may be similar to that by increasing \(P_{\rm ext}\).

\begin{figure*}[tb]
  \includegraphics{"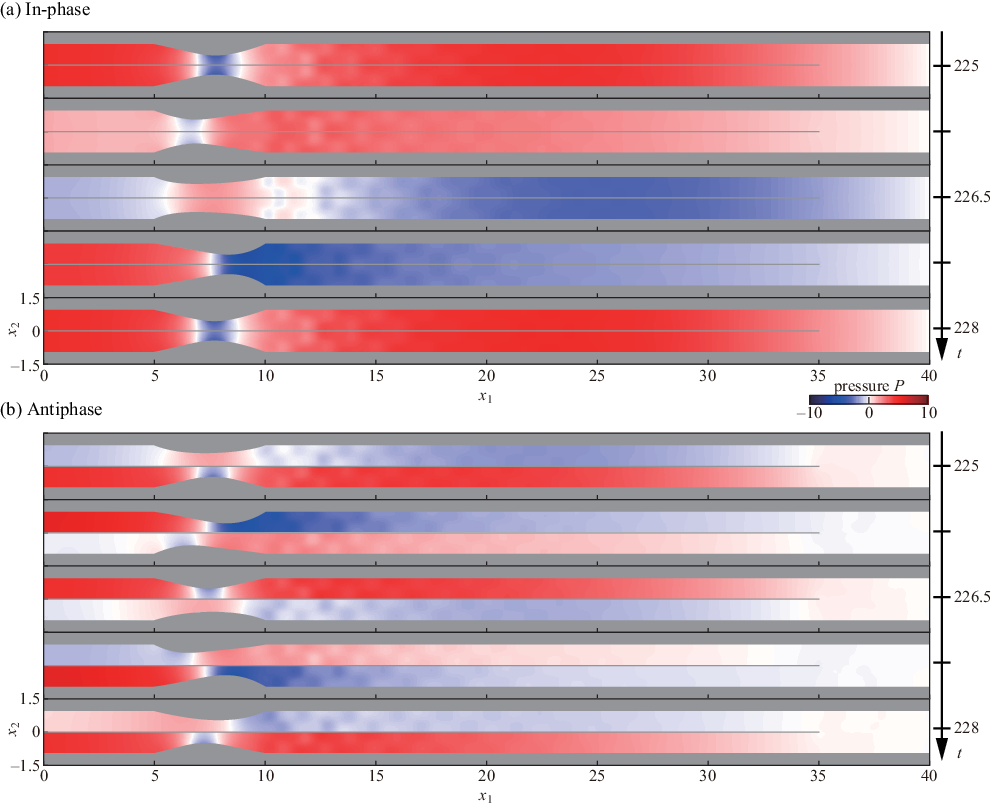"}
  \caption{
    Sequential snapshots of the pressure field \(P\) in two collapsible channels connected in parallel for \(d=25\) in the (a)~in-phase and (b)~antiphase synchronization modes after a sufficiently long time \(225\leq t\leq228\) from the state at \(t=0\), which correspond to Figs.~\ref{fig:6}(a) and \ref{fig:6}(b), respectively.
  \label{fig:8}}
\end{figure*}

The pressure fields for the parameters in Figs.~\ref{fig:6}(a) and \ref{fig:6}(b) are shown in Figs.~\ref{fig:8}(a) and \ref{fig:8}(b), respectively. In both the in-phase and antiphase synchronization modes, the pressure fields were almost homogeneous along the \(x_2\)-axis downstream of the merging point. The pressure field in the in-phase synchronization mode was almost the same as the single collapsible channel shown in Appendix~\ref{app:1}, while the pressure field downstream of the merging point in the antiphase synchronization mode did not change over time. We also performed the simulation with an additional setting of a wall at \(x_2=0\) downstream of the merging point and the result suggested that the primary interaction was the pressure field. For details, see Appendix~\ref{app:2}. Therefore, the pressure changes in the entire system due to the moving boundary significantly influenced the coupling between the oscillators, which might lead to complex synchronization phenomena with the change in the stability of the synchronization mode.

\begin{figure}[tb]
  \includegraphics{"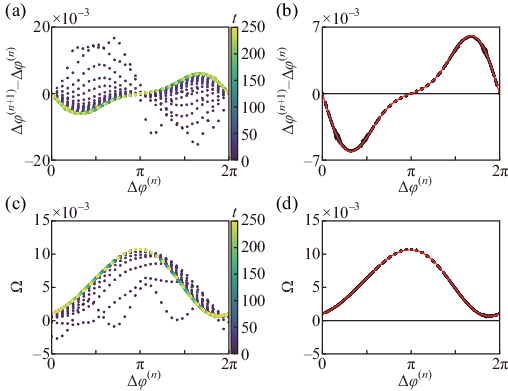"}
  \caption{
    (a),~(b)~Scatter plots of the difference \(\Delta\varphi^\qty(n+1)-\Delta\varphi^\qty(n)\) against \(\Delta\varphi^\qty(n)\) for \(d=29.5\) for the various initial conditions. We plot the entire time series in panel~(a) and only the time series after some time (\(n\geq30\)) in panel~(b). (c),~(d)~Scatter plots of the phase coupling function \(\Omega\qty(\Delta\varphi^\qty(n))\) against \(\Delta\varphi^\qty(n)\) for \(d=29.5\) for the various initial conditions. We plot the entire time series (\(n\geq 0\)) in panel~(c) and only the time series after some time (\(n\geq30\)) in panel~(d). The red lines in panels~(b) and (d) indicate the fitting curve using the Fourier series up to the third order. 
  \label{fig:9}}
\end{figure}

We investigated \(d\)-dependences of the stable synchronization modes in two collapsible channels connected in parallel as shown in Fig.~\ref{fig:7}(d). Around \(d=30\), since the interaction between two collapsible channels was small and it took time to reach a converged state,
we could not clarify the stable synchronization modes. Thus we tried to clarify the stable synchronization modes by considering the phase dynamics. Here we measured \(\Delta\varphi^\qty(n)\), which is \(\varphi_{\rm U}-\varphi_{\rm L}\) at \(\varphi_{\rm L}=0\) in the \(n\)-th period for \(t\geq0\). We show the scatter plot of the time difference \(\Delta\varphi^\qty(n+1)-\Delta\varphi^\qty(n)\) against \(\Delta\varphi^\qty(n)\) for \(d=29.5\), all \(n\), and all \(\Delta\varphi^\qty(0)\) in Fig.~\ref{fig:9}(a). We also show the same plot only for \(n\geq30\) (\(t \gtrapprox 90\)) in Fig.~\ref{fig:9}(b). We could observe the convergence to a single curve. This curve should indicate the stability of the phase difference.  \(\Delta\varphi^\qty(n)=0\) and \(\pi\) are the fixed points because the time evolution of the phase difference \(\Delta\varphi^\qty(n+1)-\Delta\varphi^\qty(n)\) is zero. The phase difference \(\Delta\varphi^\qty(n)\) in the range of \(0<\Delta\varphi^\qty(n)<\pi\) should decrease and converge to 0 because the time evolution of the phase difference \(\Delta\varphi^\qty(n+1)-\Delta\varphi^\qty(n)\) is negative. \(\Delta\varphi^\qty(n)\) in the range of \(\pi<\Delta\varphi^\qty(n)<2\pi\) should increase and converge to \(2\pi\) (equivalent to 0) because \(\Delta\varphi^\qty(n+1)-\Delta\varphi^\qty(n)\) is positive. Therefore, the in-phase mode \((\Delta\varphi^\qty(n)=0)\) should be stable while the antiphase mode \((\Delta\varphi^\qty(n)=\pi)\) should be unstable for \(d=29.5\).

The convergence to a single curve as shown in Fig.~\ref{fig:9}(b) means that the interaction strength is small and the phase dynamics depend only on the phase. In such a case, the phase of each oscillator is governed by the phase dynamics,
\begin{align}
  &\dv{\varphi_{\rm U}}{t}=\omega+\Omega(-\Delta\varphi),\\
  &\dv{\varphi_{\rm L}}{t}=\omega+\Omega(\Delta\varphi),
\end{align}
where \(\omega\) is the intrinsic frequency and \(\Omega(\cdot)\) is the phase-coupling function. From these equations, we can obtain the equation governing the phase difference,
\begin{align}
  \dv{\Delta\varphi}{t}=\Omega(-\Delta\varphi)-\Omega(\Delta\varphi)\label{eq:5}.
\end{align}
Assuming that \(\max_{\Delta\varphi}|\Omega(\Delta\varphi)|\ll\omega\), we approximately obtain the discrete equation governing the phase-difference as,
\begin{align}
  \Delta\varphi^\qty(n+1)-\Delta\varphi^\qty(n)=\frac{2\pi}{\omega}\qty(\Omega(-\Delta\varphi^\qty(n))-\Omega(\Delta\varphi^\qty(n))).
\end{align}
The curve for \(d=29.5\) shown in Fig.~\ref{fig:9}(b) should correspond to this discrete equation governing the phase difference. Since \(2\pi/(\omega+\Omega(-\Delta\varphi^\qty(n)))\) is almost equal to the time duration between \(\varphi_{\rm U}=0\) and \(\varphi_{\rm U}=2\pi\), we measured the time duration and obtained the phase coupled function \(\Omega(\Delta\varphi^\qty(n))\). In Figs.~\ref{fig:9}(c) and \ref{fig:9}(d), we show the scatter plot of \(\Omega(\Delta\varphi^\qty(n))\) against \(\Delta\varphi^\qty(n)\) for the same condition as in Figs.~\ref{fig:9}(a) and \ref{fig:9}(b). We could also observe the convergence to a single curve, which should correspond to the phase-coupling function. Figure~\ref{fig:9}(d) indicates that \(\Omega(\Delta\varphi^\qty(n))\) holds the assumption, \(\max_{\Delta\varphi}|\Omega(-\Delta\varphi)|\ll\omega\), where \(\omega\approx10^{0}\). We could ignore the influence of the amplitude on the stable phase difference and assume that the interaction strength is small because the magnitude of \(\qty|(\Delta\varphi^\qty(n+1)-\Delta\varphi^\qty(n))/\Delta\varphi^\qty(n)|\approx10^{-2}\) is much smaller than \(2\pi\tilde\lambda/\omega=-0.23\), where \((\Delta\varphi^\qty(n+1)-\Delta\varphi^\qty(n))/\Delta\varphi^\qty(n)\) is the damping rate per period for the phase difference around \(\Delta\varphi^\qty(n)=0\), \(2\pi\tilde\lambda/\omega\) is the damping rate per period for the amplitude, and \(\tilde\lambda=-0.078\) is the second maximum Lyapunov exponent  obtained from the slope of the fitting line shown in Fig.~\ref{fig:4}. Note that the rapid convergence to the single curve (plots for \(n<30\) shown in Fig.~9(a)) should correspond to the amplitude convergence. Therefore, the interaction strength was small and the phase equation determined the stable phase difference for \(d=29.5\), i.e., the in-phase mode was stable whereas the antiphase mode was unstable.

\begin{figure}[tb]
  \includegraphics{"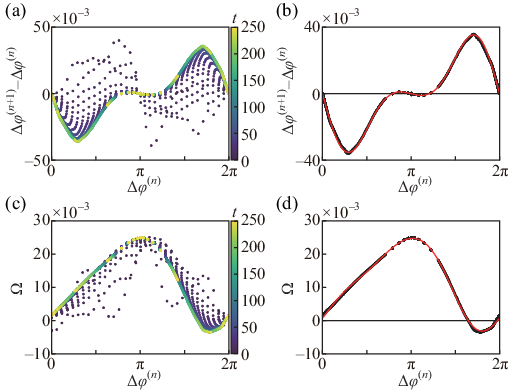"}
  \caption{
    (a),~(b)~Scatter plots of the difference \(\Delta\varphi^\qty(n+1)-\Delta\varphi^\qty(n)\) against \(\Delta\varphi^\qty(n)\) for \(d=29\) for the various initial conditions. We plot the entire time series in panel~(a) and only the time series after some time (\(n\geq30\)) in panel~(b). (c),~(d)~Scatter plots of the phase coupling function \(\Omega\qty(\Delta\varphi^\qty(n))\) against \(\Delta\varphi^\qty(n)\) for \(d=29\) for the various initial conditions. We plot the entire time series (\(n\geq 0\)) in panel~(c) and only the time series after some time (\(n\geq30\)) in panel~(d). The red lines in panels~(b) and (d) indicate the fitting curve using the Fourier series up to the third order.
  \label{fig:10}}
\end{figure}

In Fig.~\ref{fig:10}, we show the corresponding results to those shown in Fig.~\ref{fig:9} for \(d=29\). The phase equation determines the stability of the phase difference for \(d=29\) in the same way as in \(d=29.5\), and both the in-phase and antiphase modes were stable for \(d=29\). In terms of discrete dynamical systems, the antiphase mode for \(d=29\) corresponds to the stable fixed point, while that for \(d=29.5\) corresponds to the unstable fixed point. In Fig.~\ref{fig:10}(b), we observed two unstable fixed points around the stable fixed point corresponding to the antiphase mode. Therefore, the stability of the antiphase mode should be changed through the subcritical pitchfork bifurcation and there should be a bifurcation point in the range of \(29<d<29.5\).

\begin{figure}[tb]
  \includegraphics{"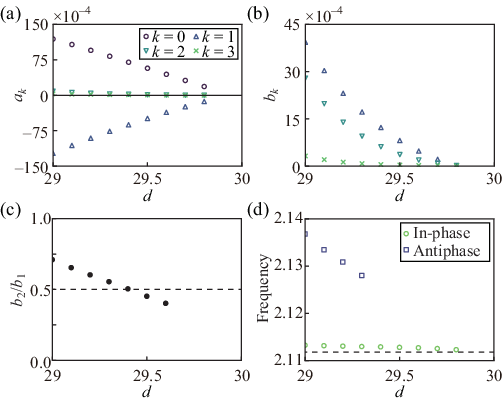"}
  \caption{
    (a)~Fourier cosine and (b)~sine coefficients of the phase coupling function \(\Omega\qty(\Delta\varphi)\) depending on \(d\). The same symbols are used for the legends in panels~(a) and (b). (c)~Ratio \(b_2/b_1\) between the first-order and second-order Fourier sine coefficients depending on \(d\). The dashed line indicates \(b_2/b_1=1/2\). (d)~Frequency depending on \(d\) for the in-phase and antiphase synchronization modes. The dashed line shows the frequency with \(d=30\).
  \label{fig:11}}
\end{figure}

To investigate the \(d\)-dependence of the stability of the antiphase mode, we considered the Fourier series of \(\Omega(\Delta\varphi)\),
\begin{align}
    \Omega(\Delta\varphi)=a_0+\sum_{k=1}^{\infty}\qty[a_k\cos(k\Delta\varphi)+b_k\sin(k\Delta\varphi)].
\end{align}
We obtained the Fourier series of \(\Omega(\Delta\varphi)\) using the data for \(n\geq30\). We show the \(d\)-dependences of the Fourier cosine and sine series of \(\Omega(\Delta\varphi)\) in Fig.~\ref{fig:11}. For all \(d\), \(\qty|a_0|\) and \(\qty|a_1|\) were significantly greater than \(\qty|a_2|\) and \(\qty|a_3|\), and \(\qty|b_1|\) and \(\qty|b_2|\) were significantly greater than \(\qty|b_3|\). Note that we also show the fitting curves using the Fourier series up to \(k=3\) for \(d=29.5\) in Figs.~\ref{fig:9}(b) and \ref{fig:9}(d) and for \(d=29\) in Figs.~\ref{fig:10}(b) and \ref{fig:10}(d), which well reproduce the phase-coupling functions.

We substituted the Fourier series of \(\Omega(\Delta\varphi)\) to Eq.~\eqref{eq:5}, as
\begin{align}
    \dv{\Delta\varphi}{t}=&-\sum_k2b_k\sin(k\Delta\varphi).\label{eq:6}
\end{align}
In the linear stability analysis, the eigenvalue \(\lambda\) around the fixed point \(\Delta\varphi=\Delta\varphi^*\) for Eq.~\eqref{eq:6} can be obtained as
\begin{align}
    \lambda=-\sum_k2kb_k\cos(k\Delta\varphi^*).
\end{align}
Only using \(b_k\) with \(k\leq2\), we obtain \(\lambda=-2b_1-4b_2\) for \(\Delta\varphi^*=0\) and \(\lambda=2b_1-4b_2\) for \(\Delta\varphi^*=\pi\). Because of \(b_1>0\) and \(b_2>0\), the in-phase mode was stable for all \(d\) as shown in Fig.~\ref{fig:11}(b). The antiphase mode is stable when \(b_2/b_1>1/2\). We show the \(d\)-dependence of \(b_2/b_1\) in Fig.~\ref{fig:11}(c). Note that we plotted only reliable data of \(b_2/b_1\) (\(d\leq29.6\)) because the small magnitudes of \(b_1\) and \(b_2\) should cause a large fitting error for \(d\geq29.7\). \(b_2/b_1\) increased as \(d\) decreased, and the antiphase mode was stable for \(d\leq29.3\) from the sign of \(b_2/b_1-1/2\). Since an unstable fixed point became stable and two unstable fixed points appeared with decreasing \(d\), we can conclude this system exhibited a subcritical pitchfork bifurcation. As a summary of analyses for the phase dynamics around \(d=30\), only the in-phase mode was stable for large \(d\), the antiphase mode became stable with a decrease in \(d\), and the stability of synchronization mode near the bifurcation point could be discussed only using the phase difference.

We show the \(d\)-dependence of the frequency for the stable synchronization mode around \(d=30\) in Fig.~\ref{fig:11}(d). For all \(d\), the frequencies both in the in-phase and antiphase modes were greater than the intrinsic frequency, and the frequency in the antiphase mode was greater than that in the in-phase mode. Using \(a_k\) only for \(k\leq1\), the frequency is expressed as \(\omega+a_0+a_1\) in the in-phase mode and \(\omega+a_0-a_1\) in the antiphase mode. Figure~\ref{fig:11}(a) indicates that the magnitudes of \(|a_0|\) and \(|a_1|\) were almost the same and the signs of them were \(a_0>0\) and \(a_1<0\) for each \(d\). These are consistent with the result that the frequency in the antiphase mode was greater than in the in-phase mode as shown in Fig.~\ref{fig:11}(d).

\begin{figure}[tb]
  \includegraphics{"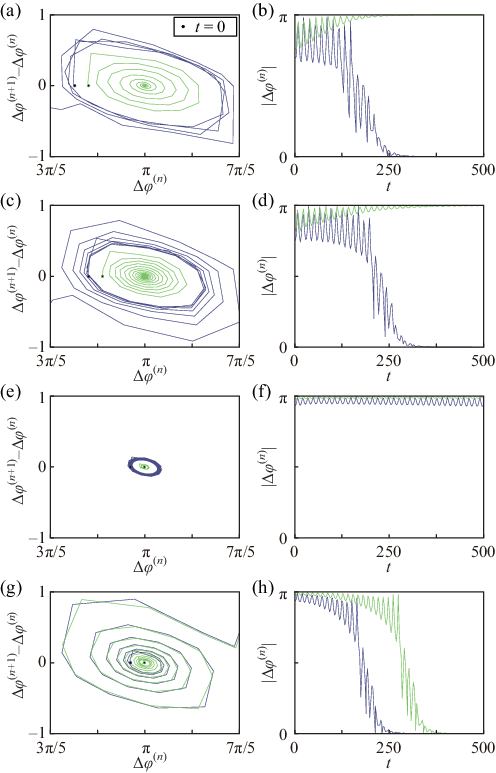"}
  \caption{
    Dynamics of \(\Delta\varphi^\qty(n)\) from two slightly different initial conditions for (a),~(b)~\(d=23.5\), (c),~(d)~\(d=23\), (e),~(f)~\(d=22.5\), and (g),~(h)~\(d=22\). (a),~(c),~(e),~(g)~Orbit in the phase plane of \(\Delta\varphi^\qty(n)\) and the time difference of \(\Delta\varphi^\qty(n)\). (b),~(d),~(f),~(h)~Time series of \(\qty|\Delta\varphi^\qty(n)|\).
  \label{fig:12}}
\end{figure}

For small \(d\), the antiphase mode became unstable with decreasing \(d\) as shown in Fig.~\ref{fig:7}(d). To investigate the bifurcation structure, we observed the phase dynamics in the same way as in the case with around \(d=30\). We show the orbit in the \(\Delta\varphi^\qty(n)\)-\((\Delta\varphi^\qty(n+1)-\Delta\varphi^\qty(n))\) plane and the time series of \(\qty|\Delta\varphi^\qty(n)|\), the absolute value of \(\Delta\varphi^\qty(n)\) (\(-\pi\leq\Delta\varphi^\qty(n)<\pi\)), from two slightly different initial conditions for \(d=23.5\), 23, 22.5, and 22 in Fig.~\ref{fig:12}. Note that we performed the numerical simulation for a time range of \(0\leq t\leq500\) to observe the convergence of the system. For \(d=23.5\) and 23, the magnitudes of \(\qty|(\Delta\varphi^\qty(n+1)-\Delta\varphi^\qty(n))/(\Delta\varphi^\qty(n)-\pi)|\approx10^{-1}\) were almost the same as \(2\pi\tilde\lambda/\omega=-0.23\), where \((\Delta\varphi^\qty(n+1)-\Delta\varphi^\qty(n))/(\Delta\varphi^\qty(n)-\pi)\) is the damping rate per period for the phase difference around \(\Delta\varphi^\qty(n)=\pi\) and \(2\pi\tilde\lambda/\omega\) is the damping rate per period for the amplitude. Therefore, the interaction strength should be large and we could not ignore the influence of the amplitude on the stability of the phase difference. Due to the strong interaction, we could not observe the convergence to a single curve but the rotating orbit in the \(\Delta\varphi^\qty(n)\)-\((\Delta\varphi^\qty(n+1)-\Delta\varphi^\qty(n))\) plane. For \(d=23.5\), the green line (where \(\Delta\varphi^\qty(0)\) was closer to \(\pi\)) converged to \(\pi\) and the purple line (where \(\Delta\varphi^\qty(0)\) was farther from \(\pi\)) diverged from \(\pi\), as shown in Figs.~\ref{fig:12}(a) and \ref{fig:12}(b). Fig.~\ref{fig:12}(a) suggests the presence of an unstable closed orbit, which was outside of the green line and inside of the purple line, and a stable fixed point at \(\Delta\varphi^\qty(n)=\pi\). For \(d=23\), the green and purple lines behaved similarly to those for \(d=23.5\), and the radius of the unstable closed orbit was smaller than for \(d=23.5\), as shown in Figs.~\ref{fig:12}(c) and \ref{fig:12}(d). For \(d=22.5\), the green and purple lines did not converge. This is probably because \(d=22.5\) is close to the bifurcation point. For \(d=22\), the fixed point at \(\Delta\varphi^\qty(n)=\pi\) was unstable as shown in Figs.~\ref{fig:12}(g) and \ref{fig:12}(h). Since a stable fixed point became unstable and an unstable closed orbit disappeared with decreasing \(d\), we conclude this system exhibited the subcritical Neimark-Sacker bifurcation\cite{Wiggins2003}. In contrast to the bifurcation point around \(d=30\), the stability of the synchronization mode near this bifurcation point could be discussed using both the phase difference and amplitude.

\section{conclusion}
We investigated the synchronization phenomena of oscillatory flows in two collapsible channels connected in parallel in the two-dimensional hydrodynamic simulation. The stable synchronization modes depended on the distance between the deformable region and the merging point; only an in-phase mode is stable for the large distance, in-phase and antiphase modes are bistable for the middle distance, and again only an in-phase mode is stable for the small distance. An antiphase mode becomes stable through the subcritical pitchfork bifurcation with decreasing distance, and the stability near this bifurcation point is discussed using only the phase difference. Further decreasing the distance, the antiphase mode becomes unstable through the subcritical Neimark-Sacker bifurcation. The stability near this bifurcation point is discussed using both the phase difference and amplitude. For all the distance, the frequency in the antiphase mode is greater than the intrinsic frequency, and the frequency in the in-phase mode is greater than in the antiphase mode. The amplitudes in in-phase and antiphase modes are greater and smaller than the intrinsic amplitude, respectively. In this system, the primary interaction may be through the pressure field. The behavior and magnitude of the interaction depend on \(d\), and thus, we observe these complex synchronization phenomena, the two synchronization mode transitions. Since the transitions between synchronization modes in our results can be described not by the phase reduction method for weakly coupled oscillators but by that for strongly coupled oscillators, our results may facilitate the development of the phase reduction method and phase analysis.

\appendix

\section{Pressure distribution in a single collapsible channel\label{app:1}}
\begin{figure*}[tb]
  \includegraphics{"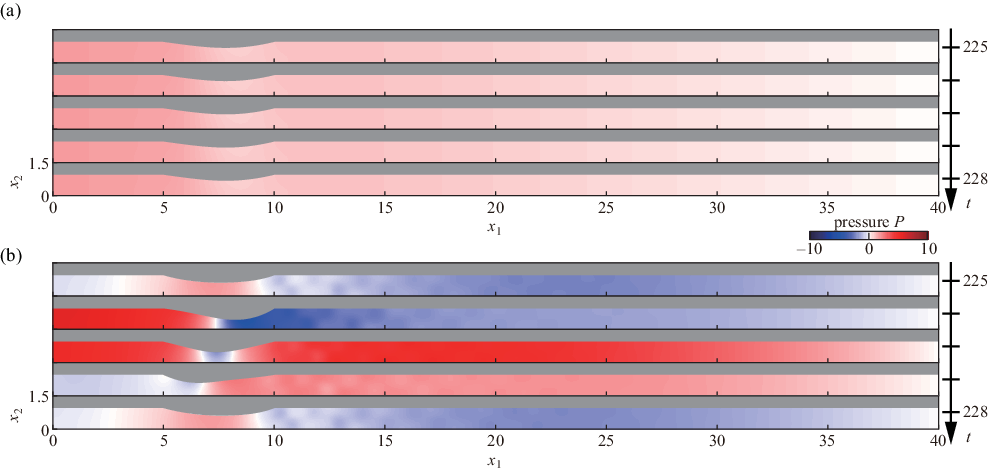"}
  \caption{
    Sequential snapshots of the pressure field \(P\) in a single collapsible channel for (a)~\(K_{\rm s}=1000, P_{\rm ext}=1.6\) and (b)~\(K_{\rm s}=1000, P_{\rm ext}=3.0\) after a sufficiently long time \(225\leq t\leq228\) from the initial state, which correspond to Figs.~\ref{fig:3}(a) and \ref{fig:3}(b), respectively.
  \label{fig:13}}
\end{figure*}

The pressure fields for the parameters in Figs.~\ref{fig:3}(a) and \ref{fig:3}(b) are shown in Figs.~\ref{fig:13}(a) and \ref{fig:13}(b), respectively. In the stationary state, the pressure field has an almost linear gradient along the \(x_1\)-axis. In the oscillatory state, the pressure changes in the entire system due to the moving boundary.

\section{Additional setting of a wall downstream of the merging point\label{app:2}}
\begin{figure*}[tb]
  \includegraphics{"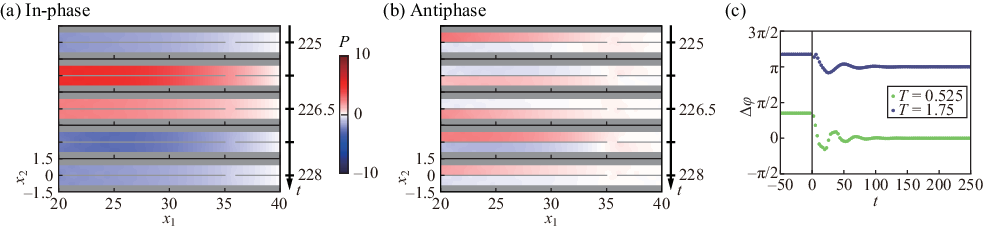"}
  \caption{
    Sequential snapshots of the pressure field \(P\) for \(d=25\) in the (a)~in-phase and (b)~antiphase modes after a sufficiently long time \(225\leq t\leq228\) from the state at \(t=0\) in the simulation with an additional setting of a wall at \(x_2=0\) downstream of the merging point. Initial conditions are different between panels~(a) and (b), where \(T=0.525\) and \(T=1.75\), respectively. (c)~Time series of the phase difference in the in-phase (green) and antiphase (purple) synchronization modes.
  \label{fig:14}}
\end{figure*}

To check whether the primary interaction was the velocity or pressure field, we performed the simulation with an additional setting of a wall at \(\{(x_1,x_2)|\, 36\leq x_1\leq40, x_2=0\}\) for \(d=25\). In this system, the upper and lower collapsible channels interact in the small region \(\{(x_1,x_2)|\, 35\leq x_1\leq36, x_2=0\}\) and the interaction through the velocity field should depend on the size of interaction region. The pressure fields downstream of the deformable region in the in-phase and antiphase modes are shown in Figs.~\ref{fig:14}(a) and \ref{fig:14}(b), respectively, and the time series of the phase difference in these modes are shown in Fig.~\ref{fig:14}(c). The additional wall hardly changed the results, and thus, the primary interaction may be not the velocity field but the pressure field.

\begin{acknowledgments}
We would like to thank Professor Makoto Iima, Professor Hiroshi Kori, and Professor Naoki Takeishi for the fruitful discussions.
This work was supported by JST, the establishment of University fellowships towards the creation of science technology innovation, Grant Number JPMJFS2107 (Y.A.), by the Cooperative Research Program of ``Network Joint Research Center for Materials and Devices.'' (Nos.~20235001 (Y.A.) and 20234003 (H.K.)), and by JSPS KAKENHI Grants Nos.~JP21K13891 (H.I.), JP20H02712, JP21H00996, and JP21H01004 (H.K.). This work was also supported by JSPS and PAN under the Japan-Poland Research Cooperative Program (No.~JPJSBP120234601 (H.K.)), and by JSPS and HAS under the Japan-Hungary Bilateral Joint Research Project (No.~JPJSBP120213801 (H.K.)).
\end{acknowledgments}

\bibliography{My_Collection}

\end{document}